\documentclass{emulateapj}
\usepackage{apjfonts}

\shorttitle{EVOLUTION OF CURRENT HELICITY}
\shortauthors{YEATES, MACKAY, \& VAN BALLEGOOIJEN}

\begin{document}

\title{Evolution and Distribution of Current Helicity in Full-Sun Simulations}

\author{A.~R.~Yeates and D.~H. Mackay}
\affil{School of Mathematics and Statistics, University of St Andrews, St Andrews, KY16 9SS, UK}
\email{anthony@mcs.st-and.ac.uk, duncan@mcs.st-and.ac.uk}
\and
\author{A.~A.~van Ballegooijen}
\affil{Harvard-Smithsonian Center for Astrophysics, 60 Garden Street, Cambridge, MA 02138, USA}
\email{vanballe@cfa.harvard.edu}

%%%%%%%%%%%%%%%%%%%%%%%%%%%%%%%%%%%%%%%%%%%%%%%%%%%%%%%%%%%%%%%%%%%%%%%%%%%%%%

\begin{abstract}
Current helicity quantifies the location of twisted and sheared non-potential structures in a magnetic field. We simulate the evolution of magnetic fields in the solar atmosphere in response to flux emergence and shearing by photospheric motions. In our global-scale simulation over many solar rotations the latitudinal distribution of current helicity develops a clear statistical pattern, matching the observed hemispheric sign at active latitudes. In agreement with observations there is significant scatter and intermixing of both signs of helicity, where we find local values of current helicity density that are much higher than those predicted by linear force-free extrapolations. Forthcoming full-disk vector magnetograms from Solar Dynamics Observatory will provide an ideal opportunity to test our theoretical results on the evolution and distribution of current helicity, both globally and in single active regions.
\end{abstract}


\section{Introduction}
%==============================================================================
The twist of the solar magnetic field plays an important role in transient phenomena such as solar flares and coronal mass ejections, and in the dynamo processes that cause the 11-year solar activity cycle. The magnetic twist can be measured in various ways. Magnetic helicity is an integral that quantifies topological complexity of field lines, such as linking, twist, or kinking (\citealp{berger1998}, \citeyear{berger1999}). For a closed magnetic system it is defined by $H_m=\int\mathbf{A}\cdot\mathbf{B}\,d^3x$, and alternative definitions have been developed for open systems \citep{berger1984,finn1985}.

In this letter we consider current helicity, which we define as
\begin{equation}
\alpha = \frac{\mathbf{j}\cdot\mathbf{B}}{B^2},
\end{equation}
where $\mathbf{B}$ is the magnetic field and $\mathbf{j}=\nabla\times\mathbf{B}$ is the current density. The quantity $\alpha$ has the advantage that it describes the \emph{local} distribution of twist and shear in the magnetic field, and that it is more readily determined from limited observational data than $H_m$ which requires global information. For a force-free field ($\mathbf{j}\times\mathbf{B}=0$) we have $\mathbf{j}=\alpha\mathbf{B}$ and $\alpha$, which may be a function of space, is a fundamental parameter that describes the torsion of the field lines around one another. Note that we shall not consider the \emph{integral} current helicity $H_c=\int\mathbf{j}\cdot\mathbf{B}\,d^3x$ because unlike $H_m$ it is not a near-conserved quantity in MHD \citep{demoulin2007}, and it does not even in general take the same sign as $H_m$ \citep[except for linear force-free fields where $\alpha$ is constant in space and $\alpha$, $H_c$, and $H_m$ all have the same sign,][]{hagyard1999}.

There are two main techniques for estimating $\alpha$ from observed vector magnetograms, which so far only cover a small region of the solar surface such as a single active region:
\begin{enumerate}
\item Compute $j_z=\partial B_y/\partial x - \partial B_x / \partial y$ and hence $\alpha_z=j_z/B_z$, which should give $\alpha$ exactly for a force-free field \citep{abramenko1996,bao1998}.
\item Compute a linear force-free extrapolation from $B_z$ and choose the overall value, $\alpha_\textrm{best}$, which best reproduces the observed $B_x$, $B_y$ distribution over the region \citep{pevtsov1995,longcope1998,zhang2006}.
\end{enumerate}
The studies by \citet{hagino2004} and \citet{burnette2004} show that both techniques are generally consistent. The key result of these observations is a robust hemispheric rule whereby the average $\alpha$ value is negative in the northern hemisphere and positive in the southern hemisphere, although there is significant scatter including a mixture of signs of $\alpha$ within single active regions. This hemispheric pattern in $\alpha$ has also been found by \citet{pevtsov2001} who reconstructed the radial and toroidal components of the global magnetic field under simplifying assumptions.

A trans-equatorial sign change in helicity is supported by numerous proxy observations such as H$\alpha$ images of active region structure \citep{hale1927}, {\it in situ} heliospheric measurements \citep{smith1993}, differential rotation \citep{berger2000}, and filament/prominence magnetic fields \citep{rust1967,martin1994}. Using newly-developed simulations of the global coronal evolution, we have recently been able to reproduce the filament hemispheric pattern including exceptions (with 96\% agreement), in a comparison with 109 observed filaments \citep{yeates2007,yeates2008}. In this letter we describe the distribution of current helicity in a 30-month simulation, which we hope to compare with new magnetic observations from the SDO (NASA Solar Dynamics Observatory) mission.

\section{Coronal Model}
%==============================================================================
\begin{figure}
  \centering
  \leavevmode
  \includegraphics*[width={\columnwidth}]{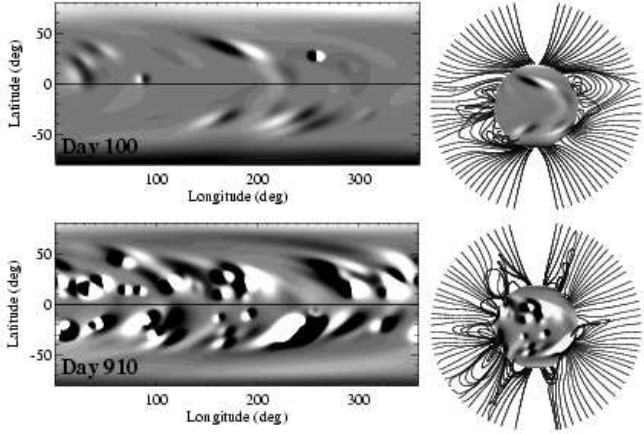}
  \hfil
  \caption{Simulated magnetic field on days 100 ({\it top}) and 910 ({\it bottom}). Left column shows radial magnetic field on solar surface (white for positive, black for negative), and right column shows selected field lines of the 3D coronal magnetic field.}
  \label{fig:field}
\end{figure}
Our simulations of the 3D coronal field evolution \citep{yeates2008} use the coupled flux transport and magnetofrictional model of \citet{vanballegooijen2000}, in a domain extending from $0^\circ$ to $360^\circ$ in longitude, $-80^\circ$ to $80^\circ$ in latitude, and $R_\odot$ to $2.5R_\odot$ in radius. The coronal magnetic field $\mathbf{B}=\nabla\times\mathbf{A}$ evolves {\it via} the non-ideal induction equation
\begin{equation}
\frac{\partial \mathbf{A}}{\partial t} = \mathbf{v}\times\mathbf{B} - \eta_c\mathbf{j},
\end{equation}
in response to flux emergence and advection by large-scale motions on the photospheric boundary. Rather than solve the full MHD system we approximate the momentum equation by the magnetofrictional method \citep{yang1986}, setting
\begin{equation}
\mathbf{v} = \frac{1}{\nu}\frac{\mathbf{j}\times\mathbf{B}}{B^2} + v_\textrm{out}(r)\hat{\mathbf{r}}.
\end{equation}
This artificial velocity ensures evolution through a sequence of near force-free states. The second term is a radial outflow imposed only near to the upper boundary, where it simulates the effect of the solar wind in opening up field lines in the radial direction \citep{mackay2006}. The diffusivity $\eta_c$ consists of a uniform background term and an enhancement in regions of strong current density $\mathbf{j}$ \citep[see][]{mackay2006}.

The photospheric boundary conditions are described in \citet{yeates2007}; the surface flux transport model includes newly emerging magnetic bipoles based on active regions observed in synoptic normal-component magnetograms from NSO, Kitt Peak. The emerging bipoles take a simple mathematical form, with properties chosen to match the location, size, tilt, and magnetic flux of the observed regions. They are inserted in 3D with a non-zero twist (magnetic helicity), chosen to match the observed sign of helicity in each hemisphere. The simulation illustrated in this letter models 30 months of continuous evolution during the rising phase of Cycle 23 (from 1997 April 9 to 1999 October 10, rotations CR1921 to CR1954). From an initial potential field extrapolation, the photospheric and coronal fields were evolved forward continuously for 914 days with 396 new bipoles inserted during this time. Two example snapshots of the simulated magnetic field are shown in Figure \ref{fig:field}.

\section{Sources of Helicity in Single Active Regions}
%==============================================================================
\begin{figure}
  \centering
  \leavevmode
  \includegraphics*[width={\columnwidth}]{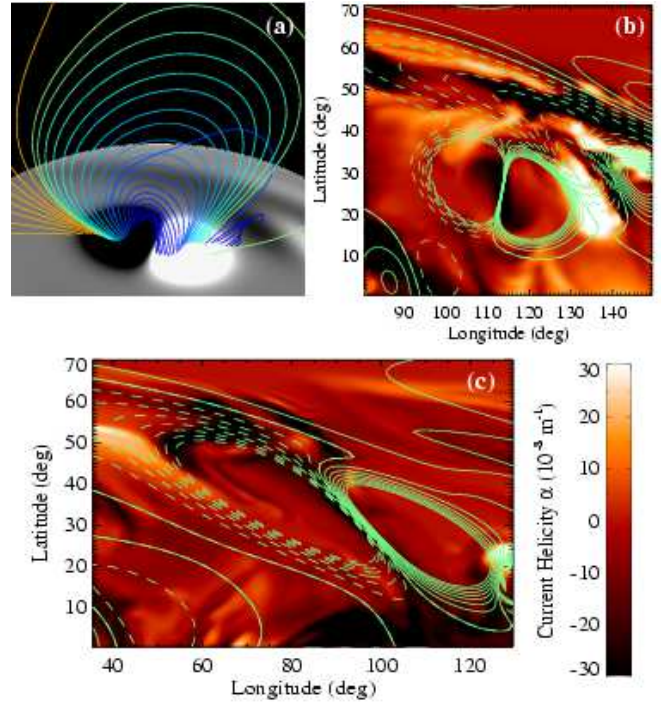}
  \hfil
  \caption{Structure of a single bipolar region, showing (a) magnetic field structure, (b) distribution of current helicity $\alpha$ on day 140, and (c) distribution of $\alpha$ on day 190. In (a) grey shading shows radial magnetic field strength on the solar surface (black negative, white positive), and coloured lines show selected coronal field lines. In (b) and (c) contours of $\alpha$ at height $14\,\textrm{Mm}$ are shown in colour scale, and green contours show strength of radial surface magnetic field (solid for positive, dashed for negative).}
  \label{fig:single}
\end{figure}
To illustrate the sources of current helicity in our simulation within an individual active region, Figure \ref{fig:single} zooms in to a bipole in the northern hemisphere which emerged on day 136 (as measured from the start of the simulation).

\begin{figure*}
  \centering
  \leavevmode
  \includegraphics*[width={2.0\columnwidth}]{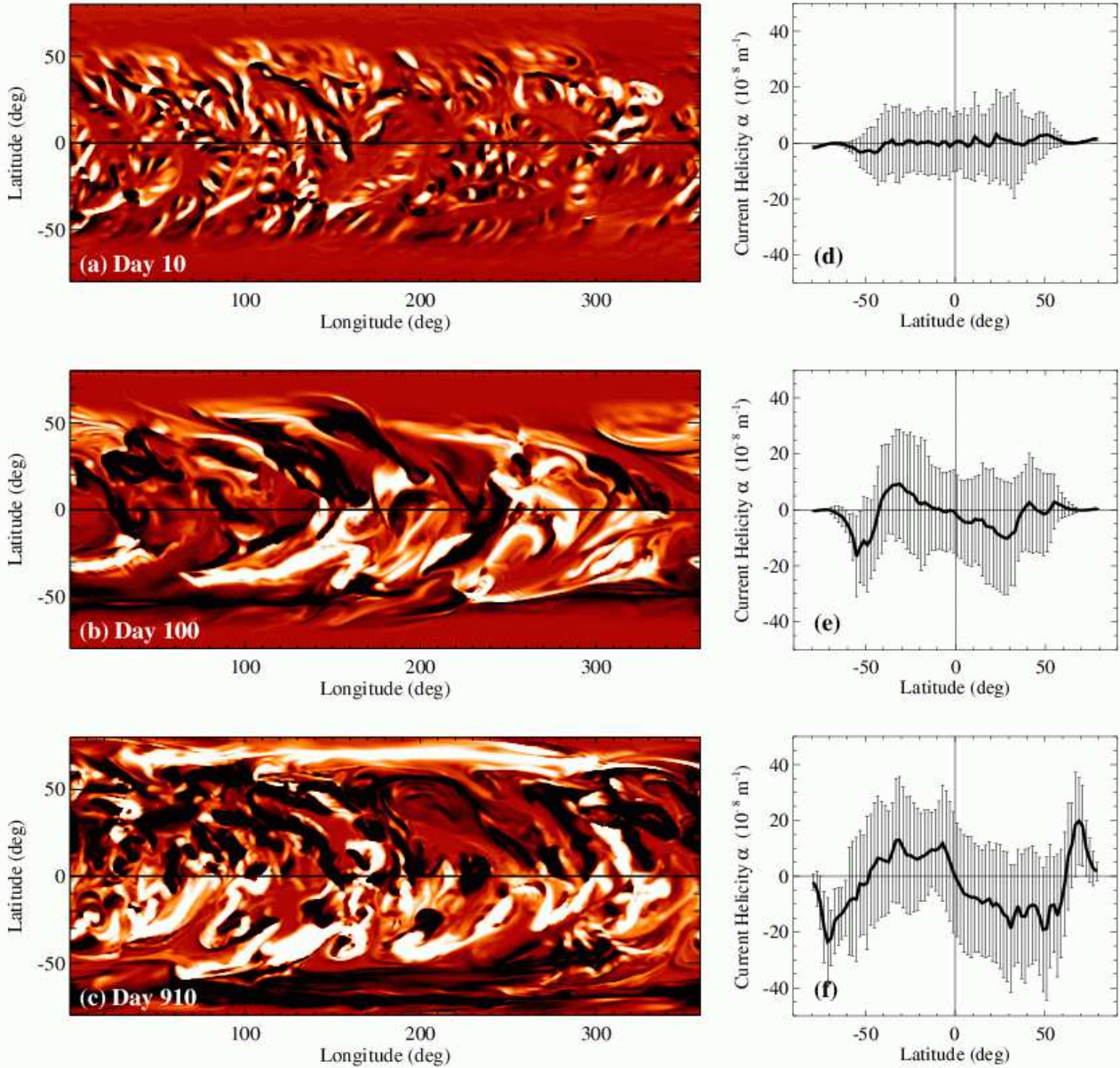}
  \hfil
  \caption{Global distribution of $\alpha$ at height $14\,\textrm{Mm}$ on days 10, 100, and 910. Left column shows contours of $\alpha$ (white for positive and black for negative, saturation level $\pm 20\times 10^{-8}\,\textrm{m}^{-1}$). Right column shows latitudinal profile, averaged over longitude in $2^\circ$ latitude bins. Error bars show one standard deviation.}
  \label{fig:global}
\end{figure*}

There are three main sources of coronal currents and helicity in our model:
\begin{enumerate}
\item The new bipoles emerge twisted. This twist is initially concentrated low down in the centre of the bipole, as seen from the field lines in Figure \ref{fig:single}(a) which are skewed as they cross the bipole's central polarity inversion line (PIL). The sigmoidal concentration of negative $\alpha$ at the centre of the bipole is clearly seen on day 140 in Figure \ref{fig:single}(b).
\item When the bipoles emerge they displace older fields and produce currents at the interface between old and new flux systems \citep[see][]{yeates2008}. In Figure \ref{fig:single}(b) this is visible at the NW edge of the new bipole where it adjoins a pre-existing bipole, and a layer of positive $\alpha$ has developed. Note that this is opposite in sign to that from the twist of the new region, as seen in Figure \ref{fig:single}(a). This corresponds to field lines that are oppositely skewed at this edge of the new bipole, as compared to those across the central PIL. This is just one example of how both signs of $\alpha$ may naturally be produced within a single active region, as found in observations.
\item Over time, surface motions shear the coronal field generating further currents. This is visible in Figure \ref{fig:single}(c), which shows the distribution of $\alpha$ for the same region on day 190, after 50 days evolution. There is a significant build-up of negative $\alpha$, particularly at the North and South ends of the bipole where helicity was initially low. This build-up is caused by differential rotation and convergence (due to supergranular diffusion).
\end{enumerate}
In addition to these sources of current helicity, it may also be locally reduced by diffusive cancellation and reconnection. Also, helicity is periodically removed through the top boundary of the domain when excessive build-up of twist leads to localised temporary losses of equilibrium, and the ejection of twisted flux ropes \citep{mackay2006}.

\section{Global Distribution of Current Helicity}
%==============================================================================

The global distribution of current helicity, $\alpha$, is shown in Figure \ref{fig:global} at days 10, 100, and 910 of the simulation. From the initial potential field on day 0 (with $\alpha=0$ everywhere), a pattern of intermixed positive and negative $\alpha$ has developed by day 10, simply due to photospheric shearing---this is before the first active region emergence. After about 100 days, a clear latitudinal trend in $\alpha$ emerges, although there is still significant local variation in both strength and sign. This pattern persists for the rest of the simulation, and up to medium heights in the 3D corona (nearer the top of the computational box high values of $\alpha$ become localized to closed field regions, with $\alpha\approx 0$ where the field is open).

In Figures \ref{fig:global}(a), (e), and (f), it can be seen how the mean $\alpha$ at low latitudes ($0^\circ$ to about $50^\circ$) develops into the observed hemispheric trend, although with considerable scatter as observed on the real Sun. However, at high latitudes the sign of $\alpha$ is reversed. These polar reversals correspond to the East-West PILs at the polar crown boundaries, and move steadily poleward through the simulation as the polar crowns reduce in size towards polar field reversal (we are approaching solar maximum). This opposite sign of $\alpha$ is caused by differential rotation of the predominantly North-South field lines at this latitude, and is a well-documented problem for theoretical models \citep{vanballegooijen1990, rust1994}. At lower latitudes, as was illustrated by Figure \ref{fig:single}(c), differential rotation of North-South PILs produces the observed hemispheric sign of helicity \citep{zirker1997}.

Figure \ref{fig:global} shows mean values of $\alpha$ at active latitudes of the order $10^{-7}\,\textrm{m}^{-1}$. The actual maximum and minimum values recorded on day 910 of the simulation were $2.24\times 10^{-6}\,\textrm{m}^{-1}$ and $-1.84\times 10^{-6}\,\textrm{m}^{-1}$. A key result of this study is that these values are much higher than those estimated from linear force-free extrapolations. Such solutions suffer a constraint on the maximum $\alpha$ in order to obtain a decay with height \citep{aulanier1998}, requiring that $\alpha<2\pi/L_x$ (the ``first resonant value''), where $L_x$ is the horizontal length of the periodic box. The linear force-free model of an observed filament by \citet{aulanier2000} has $\alpha=2.3\times 10^{-8}\,\textrm{m}^{-1}$, and for the solutions of \citet{mackay1999} this first resonant value was at $\alpha=4.24\times 10^{-8}\,\textrm{m}^{-1}$. By contrast, studies using nonlinear force-free extrapolations from vector magnetograms using the Grad-Rubin type method \citep{amari1997} find locally higher values of $\alpha$ \citep[e.g.,][]{bleybel2002}. They are also more realistic because they allow variable $\alpha$ within a single region, as in our simulations. For a particular active region, \citet{regnier2002} found maximum values of the order $10^{-6}\,\textrm{m}^{-1}$, consistent with the results of our simulations.

\section{Discussion}
%==============================================================================
In this letter we have shown how our 3D simulations of the global coronal magnetic field evolution are able to model the development and transport of current helicity, $\alpha$, over many solar rotations. We find a clear latitudinal pattern of $\alpha$ that persists throughout the simulation, although locally within single bipoles there is significant scatter and intermixing of both signs of $\alpha$, in agreement with observations. Local values may be much higher than those predicted by linear force-free extrapolations.

With existing measurements of $\alpha$ limited to vector magnetograms of individual active regions, robust observations of the latitudinal distribution of $\alpha$ await full-disk vector magnetograms. These will shortly be available from the NASA Solar Dynamics Observatory (SDO) satellite. In particular the HMI (Helioseismic and Magnetic Imager) instrument will provide synoptic full-disk vector magnetograms at 1'' resolution and approximately $90\,\textrm{s}$ cadence. This will offer an exciting opportunity to test and refine our theoretical model for the coronal magnetic field. In particular, consistent measurements over a large portion of the solar cycle will allow us to consider how the helicity distribution varies over both space and time.

Whether there is a systematic variation in the latitudinal trend of helicity over the solar cycle remains an unresolved issue \citep{sokoloff2006,pevtsov2008}, and has implications for the sub-surface origin of helicity \citep{choudhuri2004}. Indeed \citet{kleeorin2003} showed that observations of $\alpha$ in active regions provide important constraints on theories of the solar dynamo itself \citep[see also][]{sokoloff2007}. Ejection of helical fields from the corona, as included in our simulations, is also thought to play an important role in sustaining the solar cycle \citep{blackman2003}.

A particular feature of our results is the sign reversal of current helicity at the high-latitude polar crowns. This would appear to be in conflict with observations of magnetic fields in polar crown filaments, which show no such reversal in their chirality pattern \citep{rust1967,leroy1983,martin1994}. We hope to address this outstanding issue in longer simulations covering a greater portion of the solar cycle. It is not at present clear whether longer-term poleward transport of the correct sign of helicity will be enough to counteract the effect of differential rotation on the North-South oriented field lines at these latitudes. Observations of vector magnetic fields in the polar regions, such as those being made by {\it Hinode} \citep{lites2008} and soon the SDO mission, should help to constrain our models.

  %%%%%%%%%%%%%%%%%%%%%%%%%%%%%%%%%%%%%%%%%%%%%%%%%%%%%%%%%%%%%%%%%%%%%%%%%%%%
\acknowledgements
Financial support for ARY and DHM was provided by the UK STFC. DHM and AAvB would also like to thank the ISSI in Bern for support. The simulations were performed on the UKMHD parallel computer in St Andrews, funded jointly by SRIF/STFC. Synoptic magnetogram data from NSO/Kitt Peak was produced cooperatively by NSF/NOAO, NASA/GSFC, and NOAA/SEL and made publicly accessible on the World Wide Web.

  %%%%%%%%%%%%%%%%%%%%%%%%%%%%%%%%%%%%%%%%%%%%%%%%%%%%%%%%%%%%%%%%%%%%%%%%%%%%

\end{document}